\documentclass[12pt]{iopart}
\usepackage{graphicx}        
\usepackage{color,soul}
\usepackage{multirow}
\usepackage{array}
\newcommand{\ket}[1]{{\vert}#1{\rangle}}
\newcommand{\bra}[1]{{\langle}#1{\vert}}
\usepackage{float}

\begin{document}
\title{Finite-time quantum Stirling heat engine}
\author{S. Hamedani Raja}
\address{QTF Centre of Excellence, Turku Centre for Quantum Physics, Department of Physics and 
	Astronomy, University of Turku, 20014 Turku, Finland}
\author{S. Maniscalco}
\address{QTF Centre of Excellence, Turku Centre for Quantum Physics, Department of Physics and 
	Astronomy, University of Turku, 20014 Turku, Finland}
\address{ QTF Centre of Excellence, Department of Applied Physics, Aalto University, FI-00076 Aalto, Finland}
\author{G. S. Paraoanu}
\address{ QTF Centre of Excellence, Department of Applied Physics, Aalto University, FI-00076 Aalto, Finland}
\author{J. P. Pekola}
\address{ QTF Centre of Excellence, Department of Applied Physics, Aalto University, FI-00076 Aalto, Finland}
\author{N. Lo Gullo}
\address{QTF Centre of Excellence, Turku Centre for Quantum Physics, Department of Physics and Astronomy, University of Turku, 20014 Turku, Finland}

\begin{abstract}
We study the thermodynamic performance of the finite-time non-regenerative Stirling cycle used as a quantum heat engine. We consider specifically the case in which the working substance (WS) is a two-level system. The Stirling cycle is made of two isochoric transformations separated by a compression and an expansion stroke during which the working substance is in contact with a thermal reservoir. To describe these two strokes we derive a non-Markovian master equation which allows to study the dynamics of a driven open quantum system with arbitrary fast driving. We found that the finite-time dynamics and thermodynamics of the cycle depend non-trivially on the different time scales at play. In particular, driving the WS at a time scale comparable to the resonance time of the bath enhances the performance of the cycle and allows for an efficiency higher than the efficiency of the slow adiabatic cycle, but still below the Carnot bound. Interestingly, performance of the cycle is dependent on the compression and expansion speeds asymmetrically. This suggests new freedom in optimizing quantum heat engines. We further show that the maximum output power and the maximum efficiency can be achieved almost simultaneously, although the net extractable work declines by speeding up the drive.
\end{abstract}

\section{Introduction}
A flourishing research activity has developed recently around the understanding of the thermodynamic properties of quantum systems \cite{QThermoBook2018,Vinjanampathy2016,Goold2016,Silveri2017}.
Special attention has been devoted to quantum heat engines and refrigerators triggered by both new theoretical questions and technological advancements in dynamical control of microscopic systems \cite{Alicki1979,Scully2003,Niskanen2007,Quan2007, Gelbwaser2013,Kosloff2014,Alicki2015,Campisi2016,Karimi2016,Rossangel2016,Tan2017,Kosloff2017}. 
From the theoretical point of view a natural question is whether quantumness of the working substance (WS) can be exploited to achieve better performances over the classical systems. The role of quantum effects has been demonstrated for example in \cite{Rossangel2014, Uzdin2015, Jaramillo2016, Klaers2017, Deffner2018, Klatzow2019, Mukherjee2020}. It has been also ascertained that the creation of coherence between energy levels leads to inner friction and reduction of the extractable work \cite{Feldmann2000, Feldmann2002, Feldmann2003, Plastina2014, Alecce2015, Brandner2017, Pekola2019}. Thus, the sole use of a quantum WS does not in general guarantee superiority over the classical counterparts \cite{Ghosh2019}.
The study of quantum thermal machines has relied mostly on the Markovian (Lindblad) description of an open system dynamics, which guarantees non-negative entropy production rate and consistency with the second law of thermodynamics \cite{Alicki1979}. It has been shown that non-Markovian dynamics could lead to negative entropy production for the open system reduced state, however, the sum of the entropy change of the bath and the open system together is positive \cite{Marcantoni2017}. Besides these studies, non-Markovianity has been found to be influential in the performance of the quantum heat engines \cite{Mukherjee2015}, and may also enhance the output power \cite{Das2020}.

Quantum heat engines are composed of a series of strokes defined by completely-positive and trace preserving maps whose product forms the propagator over the full cycle \cite{QThermoBook2018}. Each stroke corresponds to coherent drive, dissipation to a heat bath, or simultaneous driving and dissipation. The maximum amount of extractable work produced by a cycle is obtained by ideal reversible processes which has the minimum entropy production. However, a true reversible process is infinitely slow and gives rise to zero output power. Therefore, to find a trade-off between the power and the efficiency one has to consider cycles running at finite times. Concerning the finite-time adiabatic strokes, i.e. coherent drive on the isolated WS, the shortcut-to-adiabaticity approach provides a way to mimic the adiabatic process \cite{Torrontegui2013, Odelin2019} and has recently been demonstrated experimentally with superconducting circuits \cite{Vepsalainen2019, Vepsalainen2020}. This technique has been employed, for example, to boost the performance of an Otto refrigerator \cite{Funo2019} and an Otto heat engine \cite{Abah2019}. The Otto cycle with a quantum WS has been extensively studied in the adiabatic as well as in the non-adiabatic case \cite{Alecce2015, Campisi2016, Jaramillo2016, Karimi2016, Kosloff2017, Deffner2018, Abah2019, Das2020}.

The quantum Carnot and Stirling cycles running in finite times have received less attention. The main reason is that a finite-time isothermal stroke is more tricky to study and optimize due to the simultaneous drive on the WS and its coupling to the heat bath. Usually, a slow drive whose effect falls within the validity of the adiabatic limit is assumed, allowing one to ignore the non-adiabatic effects \cite{Vacanti2014, Suri2018, Scandi2019}. This assumption is relaxed in the derivation of a time-dependent Markovian master equation to capture non-adiabatic effects but retaining the assumption that the time scales of the external drive and the ones of the coupling to the bath are still well-separated \cite{Dann2018}. Using this master equation, it has been proposed to reverse-engineer the thermalization to find a corresponding driving protocol which provides shortcut to equilibration \cite{Dann2019}. Alternatively, manipulating the coupling between the WS and the bath is also shown to speed-up isothermal strokes \cite{Pancotti2020}. In particular, the Stirling cycle has been studied in the ideal adiabatic regime \cite{Huang2014, Yin2017} and only very recently a finite-time scenario in an optomechanical implementation has been studied~\cite{Serafini2020}. There the compression and expansion strokes have been treated in the Markovian regime and the adiabatic limit and, as the authors state, a deeper investigation to include non-Markovian and out-of-equilibrium effects is still lacking. 

Here we fill this gap by studying the thermodynamics of the Stirling cycle, used as a quantum heat engine, in finite-times. A two-level system is considered as the WS and we investigate the role of different time scales involved in its dynamics. The description of the compression and expansion strokes in contact with the thermal baths requires solving the real-time dynamics of an open system beyond the adiabatic limit. In this work, we analyze the isothermal stroke in finite times without any restriction on the time scale of the drive, thus allowing for the dynamics to be non-Markovian. We note that in the non-adiabatic regime the quantum system, although in contact with a bath at fixed temperature, is brought out-of-equilibrium and its temperature is in general not defined. Hereafter by \textit{isothermal} we shall refer to the fact that the WS is in contact with a bath in equilibrium at a well defined temperature. To study the dynamics of the WS during the isotherms we derive a non-Markovian master equation using the results presented in Refs~\cite{Kofman2004, Gordon2007}. We show that the master equation contains two time-dependent parts, the rotating (R) and the counter-rotating (CR) terms. Each part also includes a Lamb shift term with the important difference that the one coming from the rotating part commutes with the non-interacting Hamiltonian whereas the second one does not. 
 
We observe that the efficiency of the cycle depends on the interplay between the driving time, the bath's correlation time and the resonance time of the hot and the cold baths. Interestingly, the efficiency exceeds that of an ideally slow cycle if we drive the qubit at a frequency comparable to the resonance frequency of the baths. The average output power also shows a similar behavior, allowing to get maximum power and maximum efficiency approximately at the same time scale for the drive. This is however not true for the net extractable work, which decreases as we speed up the drive. Our results also show that the performance of the cycle is non-trivially dependent on the individual speed in the compression and expansion strokes when the qubit is coupled to the hot and the cold bath respectively. As this dependence is in general asymmetric regarding the cold and the hot bath, it opens the possibility to optimize the performance of the cycle by choosing asymmetric compression and expansion speeds/protocols.  


The paper is organized as follows: in Section 2 we introduce the master equation,
followed by the presentation of the Stirling engine in Section 3. The calculation of work and heat for the Stirling engine is done in Section 3. Section 4 deals with the evaluation of the thermodynamic performance of the engine. Finally, section 6 is devoted to concluding remarks. 

\section{The Master Equation}
To study the dynamics of a driven WS in contact with a thermal bath we employ a non-Markovian master equation obtained by applying the approach developed in Refs. \cite{Kofman2004, Gordon2007}. There, assuming weak coupling to the baths and the Born approximation, a general time-convolutionless non-Markovian master equation is derived using the Nakajima-Zwanzig method.
Such a master equation is valid for any characteristic time scale of the drive, e.g. the period in a periodic drive or the ramping time in the case of a switching. The master equation retains both rotating and counter-rotating contributions, where the latter is specifically non-negligible at fast driving speeds. 
Here we introduce the operatorial form of the master equation and discuss its main features, leaving more details on the numerical implementation and how to recast the master equation in the adiabatic basis in appendix A. 

Let us consider a quantum system subject to an external coherent driving field and weakly coupled to a thermal bath at an inverse temperature $\beta$. The total Hamiltonian reads ($\hbar=1$)
\begin{equation}
\hat{H}(t)=\hat{H}_S(t)+\hat{H}_I(t)+\hat{H}_B,
\end{equation}
where $\hat{H}_S(t)$ and $\hat{H}_B(t)$ are the bare Hamitlonian of the open system and the bath respectively. We write the interacting Hamiltonian in the form
\begin{equation}
\hat{H}_I(t)=\hat{S}(t)\otimes \hat{B},
\end{equation}
with $\hat{S}(t)$ being a time-dependent operator acting on the open system, and $\hat{B}$ an operator acting on the bath. 
The non-Markovian master equation reads \cite{Kofman2004}
\begin{equation}
\mathcal{L}_t[\hat{\rho}(t)]=-i[\hat{H}_S(t),\hat{\rho}(t)]+\int_0^t d\tau\;\left[\Phi(t-\tau)[\hat{\tilde{S}}(t,\tau)\hat{\rho}(t),\hat{S}(t)]+{\rm h.c}\right],
\end{equation}
where 
\begin{eqnarray}\label{eq4}
&&\hat{\tilde{S}}(t,\tau)=\hat{U}(t,\tau)\hat{S}(\tau)\hat{U}(t,\tau)^{\dagger},
\\
&&\hat{U}(t,\tau)={\rm T}_{+}e^{-i\int_{\tau}^t ds \hat{H}_S(s)},
\\
&&\Phi(t)=\langle e^{i \hat{H}_B t}\hat{B} e^{-i \hat{H}_B t}\hat{B}\rangle_{\rho_B}.
\end{eqnarray}
Here $\Phi(t)$ is the correlation function of the bath and $\rho_B$ denotes the equilibrium state of the bath at an inverse temperature $\beta$. The correlation function is related to the bath's coupling spectrum $G_{\beta}(\omega)$ via the Fourier transform $G_{\beta}(\omega)=\int_{-\infty}^{+\infty} ds~\Phi(s)e^{i\omega s}$.

By decomposing the operators in the master equation w.r.t. the instantaneous eigenvectors of $\hat{H}_S(t)$, denoted by $\{\ket{\epsilon_i(t)}\}$, we get (see appendix A for more details)
\begin{equation}\label{ME}
\mathcal{L}_t[\hat{\rho}(t)]=-i\left[\hat{H}_{eff}(t),\hat{\rho}(t)\right]
+\mathcal{D}^{(R)}_t[\hat{\rho}(t)]+\mathcal{D}^{(CR)}_t[\hat{\rho}(t)].
\end{equation}
where $\hat{H}_{eff}(t)=\hat{H}_S(t)+\hat{H}^{(R)}_L(t)+\hat{H}^{(CR)}_L(t)$ with $\hat{H}_L^{(\alpha)}(t)$ with $\alpha=R,CR$ is the rotating/counter-rotating Lamb shift in the energy levels of the system generated by the coupling to the bath. Also the non-unitary dissipators $\mathcal{D}^{(R)}_t[\cdot]$ and $\mathcal{D}^{(CR)}_t[\cdot]$ account for the exchange of energy with the bath and/or decoherence.

Note that the dissipators accounting for two different baths are additive by construction if one assumes that the baths are initially uncorrelated. Naturally, the specific expressions of the different terms appearing on the r.h.s. of Eq.~(\ref{ME}) depend on the choice of the free Hamiltonian of the system and more importantly on the coupling Hamiltonian $\hat{H}_I(t)$.

\section{Quantum Stirling heat engine}
\begin{figure}
\includegraphics[width=\linewidth]{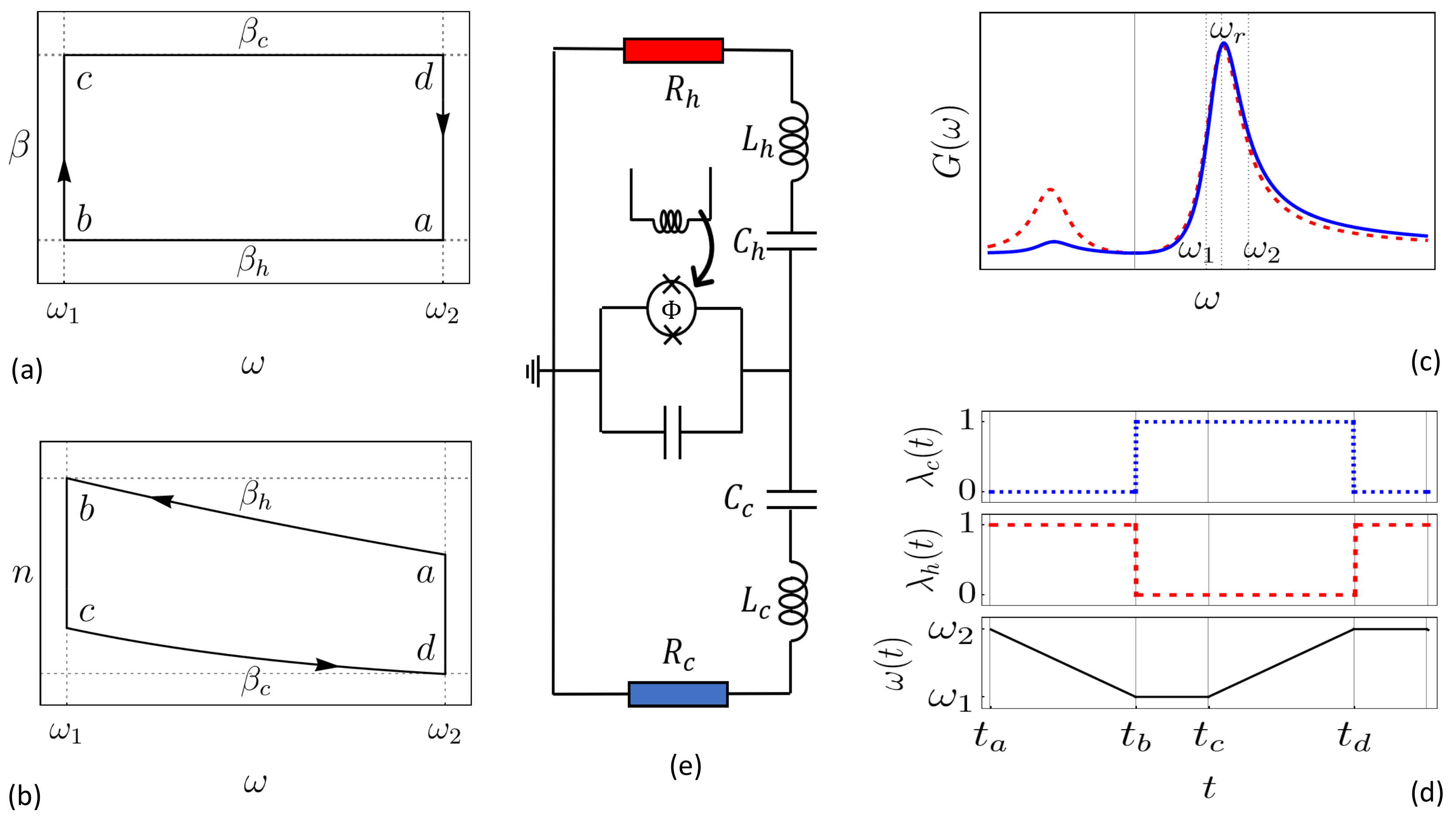}
\caption{The quantum Stirling heat cycle and its implementation using superconducting circuits. The $\beta$--$\omega$ and $n$--$\omega$ diagrams of the ideal Stirling cycle working between inverse temperatures $\beta_c$ and $\beta_h$ are respectively shown in panel (a) and panel (b), where $n(t)=\mathrm{tr}[\hat{H}_S(t)\hat{\rho}(t)]/\omega(t)$ is the instantaneous polarization of the two-level system. Panel (c) shows the coupling spectra of the two heat baths peaked at resonance frequency $\omega_r$ and the range of frequency drive of the two-level system ($[\omega_1,\omega_2]$). Panel (d) shows the piece-wise continuous coupling to the heat baths and the frequency drive of the two-level system as a function of time during a full cycle. Panel (e) shows a proposed circuit to implement the cycle using a superconducting two-level system (TLS) capacitively coupled to two RLC resonators, playing the role of the cold and hot baths. The energy levels of the two-level system is drived by tuning the external magnetic flux applied to the superconducting qubit.}
\label{Fig_cycle}
\end{figure}

The Stirling cycle is composed of two isothermal strokes and two isochoric thermalizations. Classically it is common to supplement the cycle with two extra steps which involve the interaction of WS with the so-called regenerator. The latter is typically some substance with a very high heat capacity whose task is to absorb heat from the WS during the cooling isochoric stroke and transfer this heat back to the WS during the heating isochor to improve the overall efficiency and minimize the waste heat. In this work we do not consider a regenerative setup, i.e. the WS interacts directly with the heat baths instead of the regenerator. The diagrams for temperature and polarization versus level separation for the Stirling cycle as a heat engine are depicted respectively in the panels (a) and (b) in Fig.~\ref{Fig_cycle}, where the polarization is given by $n(t)=\mathrm{tr}[\hat{H}_S(t)\hat{\rho}(t)]/\omega(t)$. Consistent with these diagrams and assuming a two-level system (TLS) as the WS, the cycle consists four strokes: 
\begin{itemize}
\item[1-] Isothermal compression, process $a\rightarrow b$, with duration $\tau_{ab}$: the level separation of the TLS reduces from $\omega_2$ to $\omega_1$ while it is coupled to the hot bath at an inverse temperature $\beta_h$. 
\item[2-] Isochoric thermalization, process $b\rightarrow c$, with duration $\tau_{bc}$: the TLS is disconnected from the hot bath and is brought to contact with the cold bath at an inverse temperature $\beta_c$, with which it thermalizes while the external drive is off. 
\item[3-] Isothermal expansion, process $c\rightarrow d$, with duration $\tau_{cd}$: the level separation of the TLS increases from $\omega_1$ back to $\omega_2$ while it is still coupled to the cold bath. 
\item[4-] Isochoric thermalization, process $d\rightarrow a$, with duration $\tau_{da}$: the TLS is disconnected from the cold bath and is brought back to contact with the hot bath. The TLS thermalizes while driving is off.
\end{itemize}
Therefore, total duration of a full cycle is $T=\tau_{ab}+\tau_{bc}+\tau_{cd}+\tau_{da}$. 

We specifically consider a setup implementable with a superconducting circuit schematically shown in the panel (e) of Fig.~\ref{Fig_cycle}. It is worth mentioning that a related design has been also put forward in \cite{Funo2019} as a possible implementation of the Otto refrigerator. The free Hamiltonian of the TLS and the TLS-bath coupling Hamiltonian are respectively denoted by $\hat{H}_S(t)$ and $\hat{H}^{(\alpha)}_I(t)$, given by
\begin{equation}\label{Hamiltonian}
\hat{H}_S(t)=\omega_0[q(t)\hat{\sigma}_z+\Delta \hat{\sigma}_x],~~
\hat{H}^{(\alpha)}_I(t)=\lambda_{\alpha}(t)\hat{\sigma}_y \otimes \hat{B}_{\alpha}.
\end{equation}
Here $\omega_0$ denotes a reference energy scale for the non-driven qubit. The operator $\hat{B}_{\alpha}$ acts on the cold/hot bath, with $\alpha=c,h$, and incorporates the coupling amplitudes between the WS and the corresponding bath. In order to realize the connection and disconnection from the baths required at steps $b$ and $d$ in the cycle, a tunable coupling element between the TLS and the resistor is required. Several types of tunable couplers have been proposed and studied, e.g. based on dressed states \cite{Paraoanu2006}, additional qubits \cite{NiskanenScience2007}, additional single Josephson junctions with current bias \cite{Blais2003}, and using a SQUID junction whose effective inductance is modulated by a bias magnetic field \cite{Mundada2019,Chen2014}. As our main motivation here is not the realization of the setup, and for the sake of simplicity, we just assume an ideal connection/disconnection protocol described by a piece-wise continuous function $\lambda_{\alpha}(t)$, as shown in the panel (d) of Fig.~\ref{Fig_cycle}.\\

As depicted in the panel (d) of Fig.~\ref{Fig_cycle}, we choose the driving protocol $q(t)$ such that the instantaneous level separation $\omega(t)=2 \omega_0\sqrt{q(t)^2+\Delta^2}$ of the TLS changes linearly with time within the interval $[\omega_1,\omega_2]$ with a given constant speed. This requirement fixes unambiguously $q(t)=\sqrt{\omega(t)^2/4-\Delta^2}$. A relevant coupling spectrum for the baths regarding the setup considered in this work is shown in the panel (c) of Fig.~\ref{Fig_cycle} and takes a specific expression given by \cite{Funo2019} 
\begin{eqnarray}
G_{\beta_i, g_i}(\omega\geq 0)&=&\frac{g_i^2}{1+f_i^2(\frac{\omega}{\omega_i}-\frac{\omega_i}{\omega})^2}\times\frac{\omega}{1-e^{-\beta_i \omega}}.
\end{eqnarray}
With $i=c,h$ denoting again the cold and hot baths, the coupling strength to the bath is described by $g_i$, the resonance frequency of the bath is denoted by $\omega_i=1/\sqrt{L_iC_i}$, and $f_i=R_i^{-1}\sqrt{L_i/C_i}$ is the quality factor of the resonators. We assume identical resonance frequency for the two baths denoted by $\omega_r$ and set the values of coupling strengths $g_c$ and $g_h$ such that the corresponding spectra have the same amplitudes at $\omega_r$ (see panel (c) of Fig.~\ref{Fig_cycle}). All the relevant physical parameters and their values used in this work are reported in Tab. 1.
\begin{table}
\small
\centering
\begin{tabular}{|c|c|c|}
\hline 
{\bf Parameter} & {\bf Definition} & {\bf Value}  \\ 
\hline 
$\tau_R$ & Relaxation time of the TLS & $1/G_{\beta_i,g_i}(\omega_r)$ \\
\hline
$\tau_B$ & Resonance time of the bath & $2\pi/\omega_r$ \\
\hline
$\tau_C$ & Correlation time of the bath & Controlled by $f$ \\
\hline   
$\beta_h$ & Inverse temperature of the hot bath & $2/\omega_0$ \\
\hline 
$\beta_c$ & Inverse temperature of the cold bath & $5/\omega_0$ \\
\hline
$(g_c,g_h)$     & Set of TLS-bath coupling amplitudes  & $g_1=(0.2,0.17)$ or $g_2=\sqrt{2}\times g_1$ \\
\hline
$\omega_r$   &  Resonance frequency of the baths &  $0.6\times \omega_0$ \\
\hline
$\omega_1$   &  Minimum frequency of the TLS &  $0.49\times\omega_0$ \\
\hline
$\omega_2$   &  Maximum frequency of the TLS &  $0.78\times\omega_0$ \\
\hline
$f$   &   Quality factor of the bath's resonators & $2$ or $3$ \\
\hline
$\tau_D$   &  Unit of driving duration & $\tau_R(g_1)$ \\
\hline
$\tau_{th}$   &   Duration of the isochoric strokes & $6\times \tau_R(g_1)$ \\
\hline
\end{tabular} 

\caption{Definitions and values of the relevant physical parameters used in this work. Note that $\omega_0$ is the reference energy scale of the TLS, with respect to which we normalize all other frequencies and time scales. ($\hbar=1$, $k_B=1$.)}
\end{table}

\begin{figure}
\includegraphics[width=\linewidth]{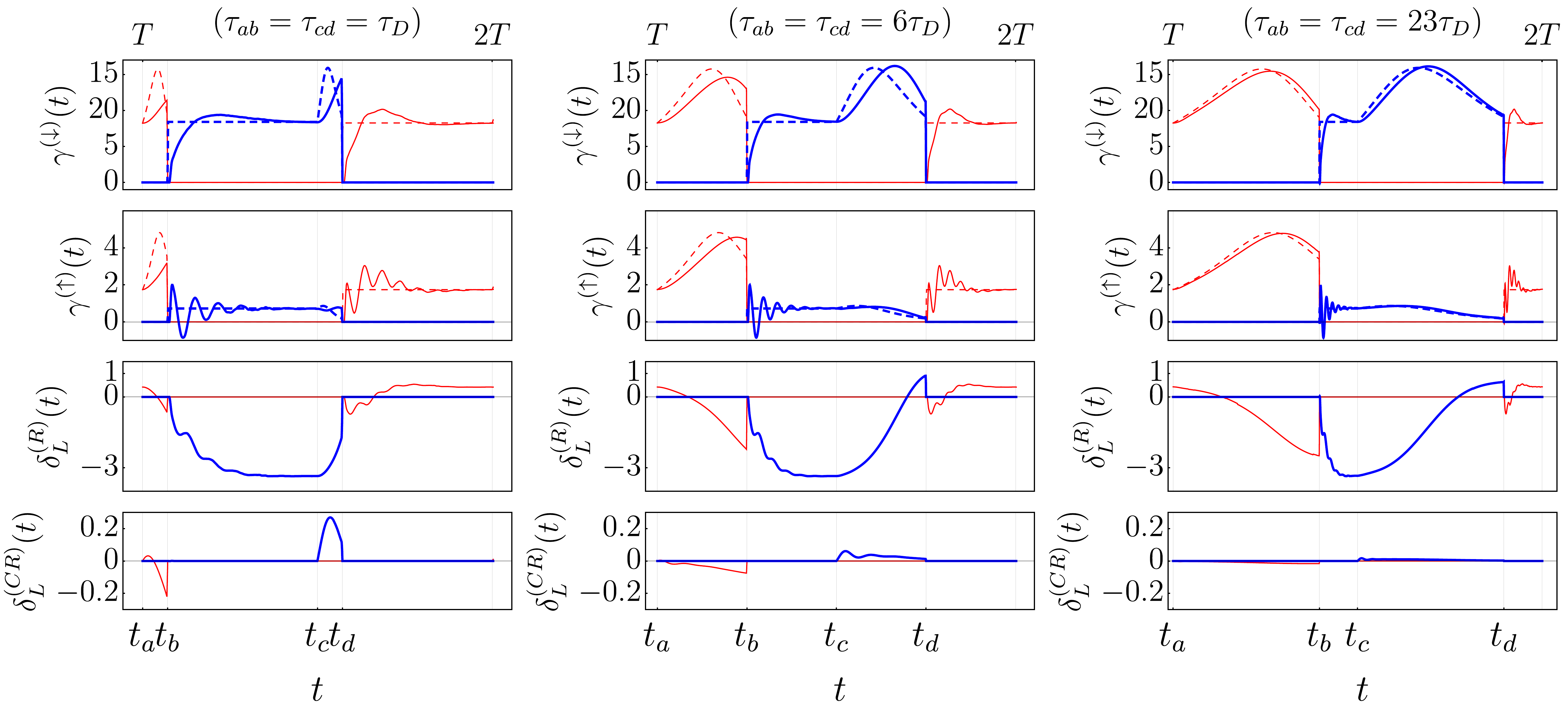}
\caption{Values of the rates of the master equation in Eq.~(\ref{MEinst}) as a function of time and for three different values of $\tau_{ab,cd}$. The two upper panels show the instantaneous transition rates between the adiabatic energy levels of the TLS plotted using the solid curves, while the dashed lines are the asymptotic Markovian limit of the rates which equal $2\pi G(\pm\omega(t))$. In the two lower panels the rotating and counter-rotating Lamb shift contributions are plotted as a function of time. Note that values of the rates are normalized by $[10^{-2}]$. In calculating the rates we have set $f=2$ and  $(g_c,g_h)=g_1$.}
\label{Fig_rates}
\end{figure}

For the specific Hamiltonian given in Eq.~(\ref{Hamiltonian}), the instantaneous energy basis of the TLS reads
\begin{eqnarray}
\ket{\epsilon_e(t)}=\cos \theta_t \ket{e}+\sin \theta_t \ket{g} \label{basis_e},
\\
\ket{\epsilon_g(t)}=\sin \theta_t \ket{e}-\cos \theta_t \ket{g} \label{basis_g},
\end{eqnarray}
with $\theta_t=(1/2)\cot^{-1}(q(t)/\Delta)$ and $\ket{e(g)}$ as the eigenbasis of $\hat{\sigma}_z$ Pauli operator. By  defining the transition operator $\hat{L}(t)=\ket{\epsilon_g(t)}\bra{\epsilon_e(t)}$ between the instantaneous energy basis of the TLS, the master equation in Eq.~(\ref{ME}) takes the explicit form
\begin{eqnarray}
\label{MEinst}
\mathcal{L}_t[\hat{\rho}(t)]=&-i\left [\Big(1+\delta^{(R)}_L(t)\Big)\hat{H}_S(t)+\delta^{(CR)}_L(t)\Big(\Delta \hat{\sigma}_z-q(t)\hat{\sigma}_x\Big),\hat{\rho}(t)\right]
\\
&+\gamma^ {(\downarrow)}(t)\Big[\hat{L}(t)\hat{\rho}(t)\hat{L}^{\dagger}(t)-\frac{1}{2}\{\hat{L}^{\dagger}(t)\hat{L}(t),\hat{\rho}(t)\}\Big] \nonumber
\\
&~~+\gamma^{(\uparrow)}(t)\Big[\hat{L}^{\dagger}(t)\hat{\rho}(t)\hat{L}(t)-\frac{1}{2}\{\hat{L}(t)\hat{L}^{\dagger}(t),\hat{\rho}(t)\}\Big]+\mathcal{D}^{(CR)}_t[\hat{\rho}(t)],\nonumber
\end{eqnarray} 
where the exact expressions for the Lamb shift terms $\delta_L^{(i)}(t)$ are given in appendix (A.2). An expression for the counter-rotating dissipator $\mathcal{D}^{(CR)}_t[\cdot]$ is, however, too cumbersome to be included here. An interesting feature of the Lamb shifts is that while the Lamb shift contribution due to the rotating term is proportional to $\hat{H}_S(t)$, the counter-rotating one does not commute with $\hat{H}_S(t)$. Temporal behavior of the rates involved in the generator $\mathcal{L}_t$ is shown in Fig.~\ref{Fig_rates} for different compression and expansion speeds during the isothermal branches. In the rest of the paper, we consider a scale for the driving duration denoted by $\tau_D$. Recalling that the amplitude of the spectra of the cold and hot baths are set to be identical at the resonance frequency $\omega_r$, the value of $\tau_D$ is fixed to the relaxation time of the TLS when the coupling amplitudes are $(g_c=0.2,g_h=0.17)$, thus $\tau_D:=\tau_R(g_1)$. In addition, duration of the isochoric branches are always fixed at $\tau_{th}=6\times\tau_R(g_1)$. 

By neglecting the counter-rotating terms in Eq.~(\ref{MEinst}) we get a time-dependent master equation in the Linblad form, which we describe by $\mathcal{L}^{(R)}_t[\hat{\rho}(t)]$. Consider this generator at a given fixed time $t=\tau$ denoted by $\mathcal{L}^{(R)}_{\tau}$, which means all the rates, Hamiltonian, and jump operators are set to their configuration at $t=\tau$ and remain unchanged for $t>\tau$. We define the invariant state of this generator by $\hat{\rho}_{eq}^{(R)}(\tau)$, such that $\mathcal{L}^{(R)}_{\tau}[\hat{\rho}_{eq}^{(R)}(\tau)]=0$. It is straightforward to check that the invariant state is given by
\begin{equation}
\hat{\rho}_{eq}^{(R)}(\tau)=\Gamma(\tau)^{-1}\left[\gamma^{(\uparrow)}(\tau)\ket{\epsilon_e(\tau)}\bra{\epsilon_e(\tau)}+\gamma^{(\downarrow)}(\tau)\ket{\epsilon_g(\tau)}\bra{\epsilon_g(\tau)}\right],
\end{equation}
with $\Gamma(\tau)=\gamma^{(\uparrow)}(\tau)+\gamma^{(\downarrow)}(\tau)$. We stress that due to the explicit time dependency of the decay rates, $\hat{\rho}_{eq}^{(R)}(\tau)$ is not necessarily identical to a Gibbs state at the same temperature of the heat bath. As depicted in the two upper panels of Fig.~\ref{Fig_rates}, it is only for the asymptotic slow driving (adiabatic limit) that the decay rates $\gamma^{(\uparrow)}(\tau)$ and $\gamma^{(\downarrow)}(\tau)$ approach to their Markovian limits $2\pi G_{\beta}(\pm\omega(\tau))$ \cite{Gordon2007} and one consequently gets the equilibrium state $\hat{\rho}_{eq}(\beta,\tau)=\exp(-\hat{H}_S(\tau) \beta)/\mathrm{tr}[\exp(-\hat{H}_S(\tau) \beta)]$, where $\beta$ is the inverse temperature of the bath with whom the TLS interacts. Asymptotic state of the full generator $\mathcal{L}_t$ which includes the counter-rotating terms is, however, more complicated and does not depend solely on the rates $\gamma^{(\uparrow /\downarrow)}(t)$, especially for fast drives. Consider the full generator at a given fixed time $t=\tau$ denoted by $\mathcal{L}_{\tau}$. We define its asymptotic state formally by $\hat{\rho}_{\tau}^*=\lim_{t\rightarrow \infty}\left[\exp( t \mathcal{L}_{\tau})\hat{\rho_i}\right]$, where $\hat{\rho}_i$ is some initial input state. We now proceed to utilize these tools to characterize the Stirling cycle used as a heat engine.

\begin{figure}
\centering
\includegraphics[width=0.75\linewidth]{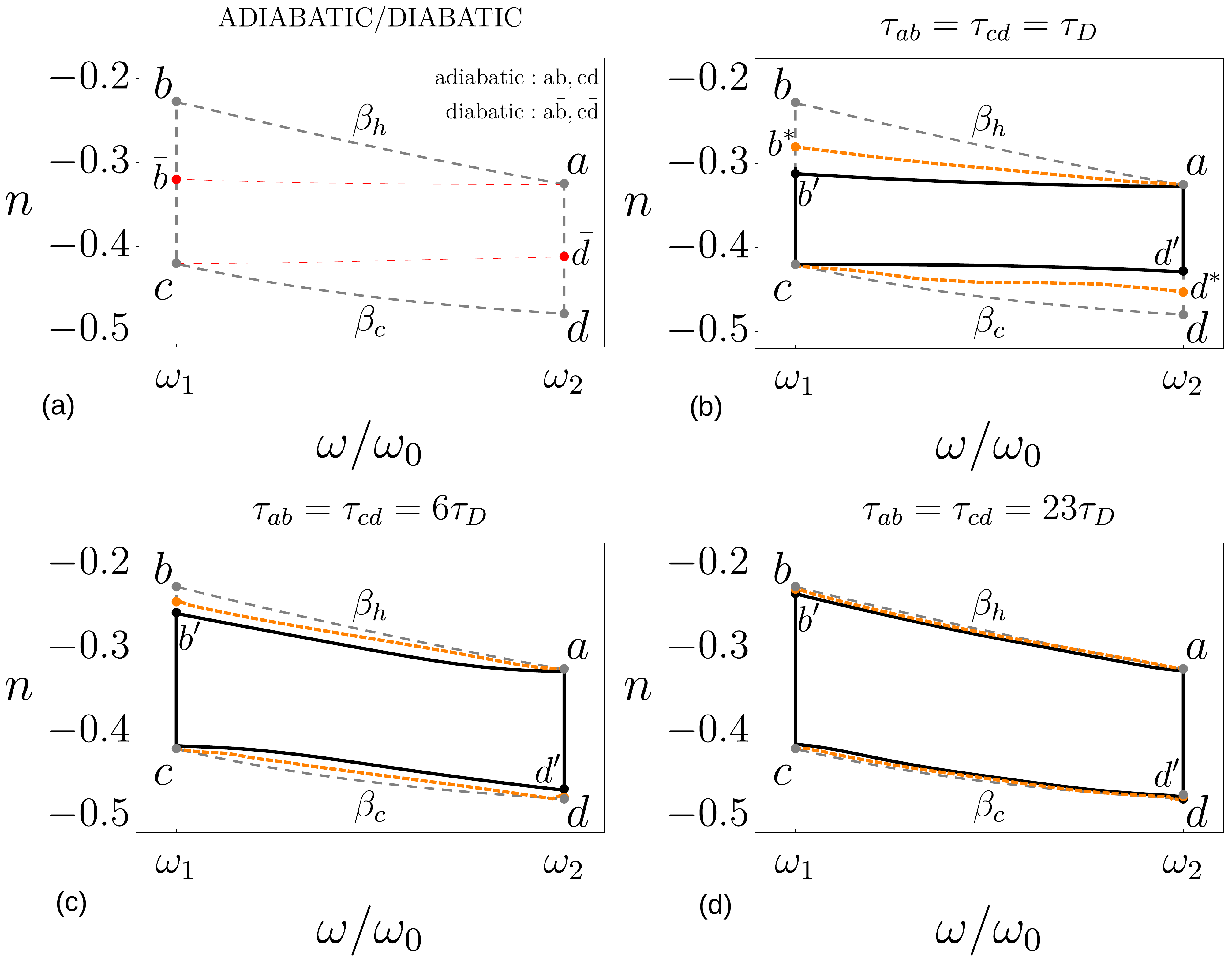}
\caption{Polarization--frequency diagram of the Stirling cycle. In panel (a) an ideal slow cycle is plotted using the dashed grey lines which follows the adiabatic trajectories $ab$ and $cd$, whereas the fast diabatic trajectories are depicted by $a\bar{b}$ and $c\bar{d}$ in dashed red. In the rest of the panels $ab'$ and $cd'$ denote the actual trajectories with finite time compression and expansion and considering three different duration. Also, $ab^*$ and $cd^*$ trajectories plotted in dashed orange denote the asymptotic steady state of the dynamics.}
\label{Fig_diagSym}
\end{figure}
We set $t_a=T$ and exclude the first cycle $0\leq t<T$ to guarantee that the rates and state of the TLS all reset to their initial values at $t=2T$, i.e. $\hat{\rho}(2T)=\hat{\rho}(T)$ and $\mathcal{L}_{2T}=\mathcal{L}_{T}$. Moreover, the duration of isochoric strokes are set sufficiently large ($\tau_{bc}=\tau_{da}=6\times\tau_R(g_1)$) such that the TLS can reach its asymptotic equilibrium states at the end of $b\rightarrow c$ and $d\rightarrow a$ branches. Accordingly, the two points $a$ and $c$ are always fixed in our analysis, as shown in the panel (a) of Fig.~\ref{Fig_diagSym}. Nonetheless, we consider arbitrary duration for the isothermal strokes. If the driving is sufficiently slow, the WS remains in an instantaneous equilibrium state with the bath during the whole process. This ideal case corresponds to the $ab$ and $cd$ trajectories in the panel (a) of Fig.~\ref{Fig_diagSym}. However, a faster drive kicks the WS out of the manifold of equilibrium states and, consequently, its trajectory deviates from the ideal isothermal(adiabatic) ones. The opposite regime is when the drive is so fast that the dynamics of the WS is essentially diabatic and its state remains unchanged during the process. Therefore, at the end of the diabatic process we end up at the point $\bar{b}(\bar{d})$, instead of adiabatic points $b(d)$. Let us now define the actual target points of the WS at the end of the compression and expansion processes with some arbitrary speeds, respectively by $b'$ and $d'$. The corresponding points of the asymptotic (equilibrium) states $\hat{\rho}_{t_b}^*$ and $\hat{\rho}_{t_d}^*$ of the WS at the end of the processes are also denoted by $b^*$ and $d^*$. Therefore, as it is shown in the panels (b), (c), and (d) of Fig.~\ref{Fig_diagSym}, by increasing the speed of driving one changes the trajectory of the WS within the two areas $a\bar{b}b$ and $c\bar{d}d$, moving from the ideal adiabatic trajectories towards the diabatic ones. Within this general picture, we now study in detail how different speeds  affect the thermodynamic performance of the Stirling heat engine.

\section{Work and heat}
Studying the performance of the heat engine requires to calculate the work done  as well as the energy exchanged with the baths during each stroke of a full cycle. 
In making the separation between work and heat it is important to include the Lamb shift in an effective Hamiltonian of the system \cite{Alipour2016}, which reads
\begin{equation}
\hat{H}_{eff}(t)=\hat{H}_S(t)+\sum_{\alpha=c,h}\left [\delta^{(R,\alpha)}_L(t)\hat{H}_S(t)+\delta^{(CR,\alpha)}_L(t)\big(\Delta \hat{\sigma}_z-q(t)\hat{\sigma}_x\big)\right],
\end{equation}
where the summation is over the terms corresponding to the hot and the cold baths. Likewise, the full dissipator acting on the WS has two parts each corresponding to one of the baths:
\begin{equation}
\mathcal{D}_t[\cdot]=\sum_{c,h}\big(\mathcal{D}^{(R,\alpha)}_t[\cdot]+\mathcal{D}^{(CR,\alpha)}_t[\cdot]\big),
\end{equation} 
Having the Hamiltonian and the dissipator of the dynamics, we can calculate the average of work and heat transferred. For the average work done on the WS in the time interval $[t_1,t_2]$ one has
\begin{equation}
\langle W(t_1,t_2)\rangle=\int_{t_1}^{t_2}ds~\mathrm{tr}\left[\left(\frac{d}{dt}\hat{H}_{eff}(t)\vert_{t=s}\right)\hat{\rho}(s)\right],
\end{equation}
which relates to the average output power $P(t_1,t_2)$ during this time interval via 
\begin{equation}
P(t_1,t_2)=(t_2-t_1)^{-1}\langle W(t_1,t_2)\rangle.
\end{equation}
The fact that the Lamb shift enters the definition of the work has important consequences. Specifically, if the Lamb shifts vary in time (which indeed happens here due to presence of a memory kernel in the master equation) it is possible to have non-zero work even when the external drive is off. Furthermore, the average heat transferred into the WS in the time interval $[t_1,t_2]$ is given by 
\begin{equation}
\langle Q(t_1,t_2)\rangle=\int_{t_1}^{t_2}ds\mathrm{Tr}\big[\hat{H}_{eff}(s)\mathcal{D}_s[\hat{\rho}(s)]\big].
\end{equation} 
By denoting $\langle W\rangle_{net}$ as the average net extractable work during a full cycle, according to the first law of thermodynamics one has $\langle W\rangle_{net}=-\langle Q \rangle_{net} $, where $\langle Q \rangle_{net}$ is the net average heat transferred. Consider the net positive heat transferred into the WS labeled by $\langle Q\rangle_h$, then the efficiency of the cycle is determined by 
\begin{equation}
\eta=\frac{\langle W\rangle_{net}}{\langle Q\rangle_h}.\\
\end{equation}

Let us recall that in a regenerative classical Stirling heat engine the heat transferred during the isochoric branch $d\rightarrow a$ is not included in calculating the efficiency, since the regenerator is an internal component of the engine. However, here we let the WS interact directly with the hot bath during the stroke. Therefore, the net positive heat transferred into the WS has contributions both from the $a\rightarrow b$ and $d \rightarrow a$ branches. Moreover, in the classical regenerative Stirling heat engine, the regenerator helps to minimize the wasted heat and increase the efficiency closer to the Carnot bound: $\eta_C=1-\beta_h/\beta_c$. As we do not resort to regeneration, we expect the efficiency to be well-below the Carnot bound $\eta_C=0.6$ (considering $\beta_c=5$ and $\beta_h=2$).

Before presenting our numerical results, we note that by considering the bare Hamiltonian $\hat{H}_S(t)$ (excluding the Lamb shifts) one can provide analytic expressions of the efficiency regarding four limiting cases. As depicted in the panel (a) of Fig.~\ref{Fig_diagSym}, these case are: $(abcda)$ trajectory corresponding to the ideal adiabatic processes - $(a\bar{b}c\bar{d}a)$ trajectory which follows the diabatic passage - $(abc\bar{d}a)$ trajectory corresponding to the adiabatic compression and diabatic expansion processes - and finally $(abc\bar{d}a)$ trajectory which follows the diabatic compression and adiabatic expansion processes. We call these cases respectively $(ss),~(ff),~(sf),~(fs)$, with $s$ and $f$ denoting \textit{slow} and \textit{fast}, respectively. The analytic analysis of efficiency for these cases is presented in Appendix B. Our aim is to examine the performance of the heat cycle for the situations between these four limiting cases and by considering the real-time evolution of the WS in finite times.

\section{Thermodynamic performances}
\begin{figure}
\centering
\includegraphics[width=\linewidth]{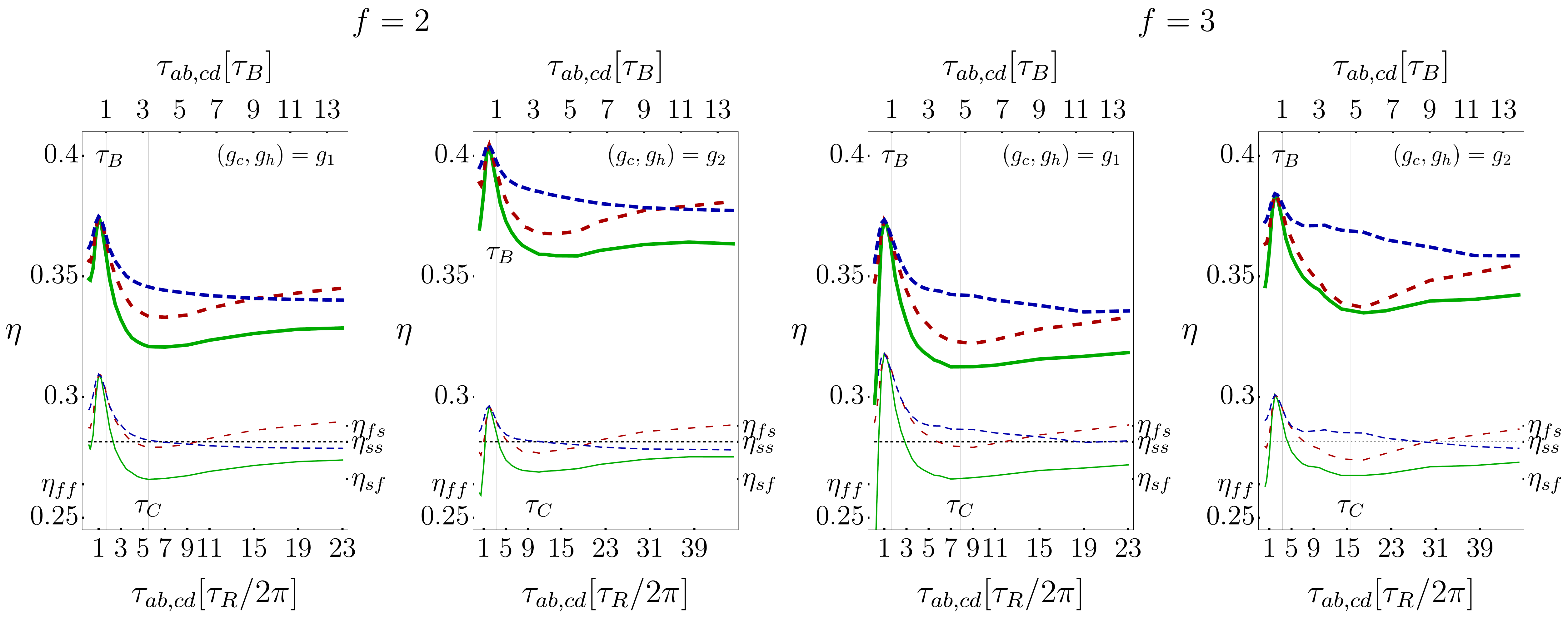}
\includegraphics[width=\linewidth]{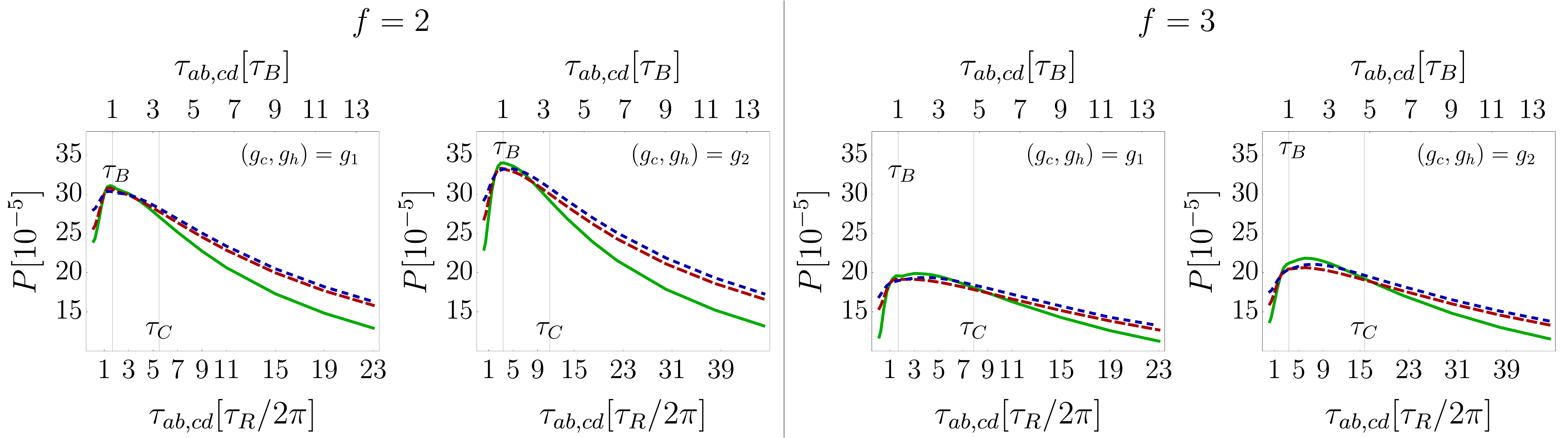}
\caption{Efficiency and output power as a function of $\tau_{ab,cd}$, respectively plotted in the upper and lower panels for different values of $f$ and $(g_c,g_h)$. Regarding the efficiency, the solid curves correspond to the symmetric driving with $\tau_{ab}=\tau_{cd}$. The asymmetric cases are plotted with the large dashing red and small dashing blue, corresponding respectively to (1): when $\tau_{ab}= \tau_D$ is fixed and $\tau_{cd}$ changes and (2): when $\tau_{cd}= \tau_D$ is fixed and $\tau_{ab}$ changes. In the upper panel, the thick curves indicate the efficiency calculated w.r.t. the effective Hamiltonian and the thin lines correspond to the calculations considering the bare Hamiltonian. Note that, however, the power is calculated only w.r.t. the bare Hamiltonian. The efficiency of the asymptotic cases discussed in the appendix B are also marked, specifically the efficiency of the ideally slow cycle $(ss)$ is plotted using the dotted black line.}
\label{Fig_Eff_Power}
\end{figure}
We first consider the situation in which the speed of compression and expansion processes are identical, i.e. $\tau_{ab}=\tau_{cd}$, which corresponds to the situations shown in the panels (b), (c), and (d) of Fig.~\ref{Fig_diagSym}. Efficiency of the cycle is plotted as a function of $\tau_{ab,cd}$ using the solid curves in the upper panel of Fig.~\ref{Fig_Eff_Power}. We have plotted the efficiency calculated using both the effective Hamiltonian $\hat{H}_{eff}$ (thick green curves) and the bare Hamiltonian $\hat{H}_S$ (thin green curves). The most striking observation is a peak in the efficiency at some values of $\tau_{ab(cd)}$ which exceeds the efficiency of the adiabatic cycle shown by the dotted black line. To realize the relation between the observed efficiency enhancement and different physical time scales involved in the dynamics of the WS, to say relaxation time $\tau_R$, bath correlation time $\tau_C$, and bath resonance time scale $\tau_B$, we have plotted the efficiency for four different cases. These are considering two different values of the relaxation time, $\tau_R(g_1)$ and $\tau_R(g_2)=\tau_R(g_1)/2$, and two different values of the bath correlation time, $\tau_C(f=2)$ and $\tau_C(f=3)=1.43\times\tau_C(f=2)$. Moreover, we set the value of $\tau_B$ fixed for all the four mentioned cases. Looking at Fig.~\ref{Fig_Eff_Power}, it is clear that the relevant parameter for the observed peak in the efficiency is the bath resonance time $\tau_B$, such that when $\tau_{ab(cd)}$ are close to $\tau_B$ we observe the enhancement in the efficiency. On the contrary, it is clear that when the time scale of the drive is close to the bath correlation time $\tau_C$ the efficiency decreases. The output power of the cycle for the same settings is also plotted as a function of $\tau_{ab,cd}$ in the lower panel of Fig.~\ref{Fig_Eff_Power}. We note that since we are interested in the work done by the external drive, the output power is only calculated with respect to the bare Hamiltonian. Interestingly, the average output power benefits from enhancement when $\tau_{ab,cd}\simeq \tau_B$ as well. However, the peak in the power is happening at a slightly larger time scale than those for the efficiency. As expected, the output power decreases by increasing $\tau_{ab,cd}$. The same behavior also holds for very short time scales, when the extractable work diminishes at ultra-fast driving due to an increase in the irreversibility.

Studying the energy flow from or into the WS is essential to comprehend the observed boost in  efficiency and power. Due to the limited space and without the loss of generality, we present the energetic results only for the case with $f=2$ and $(g_c,g_h)=g_1$.
\begin{figure}
\centering
\includegraphics[width=0.8\linewidth]{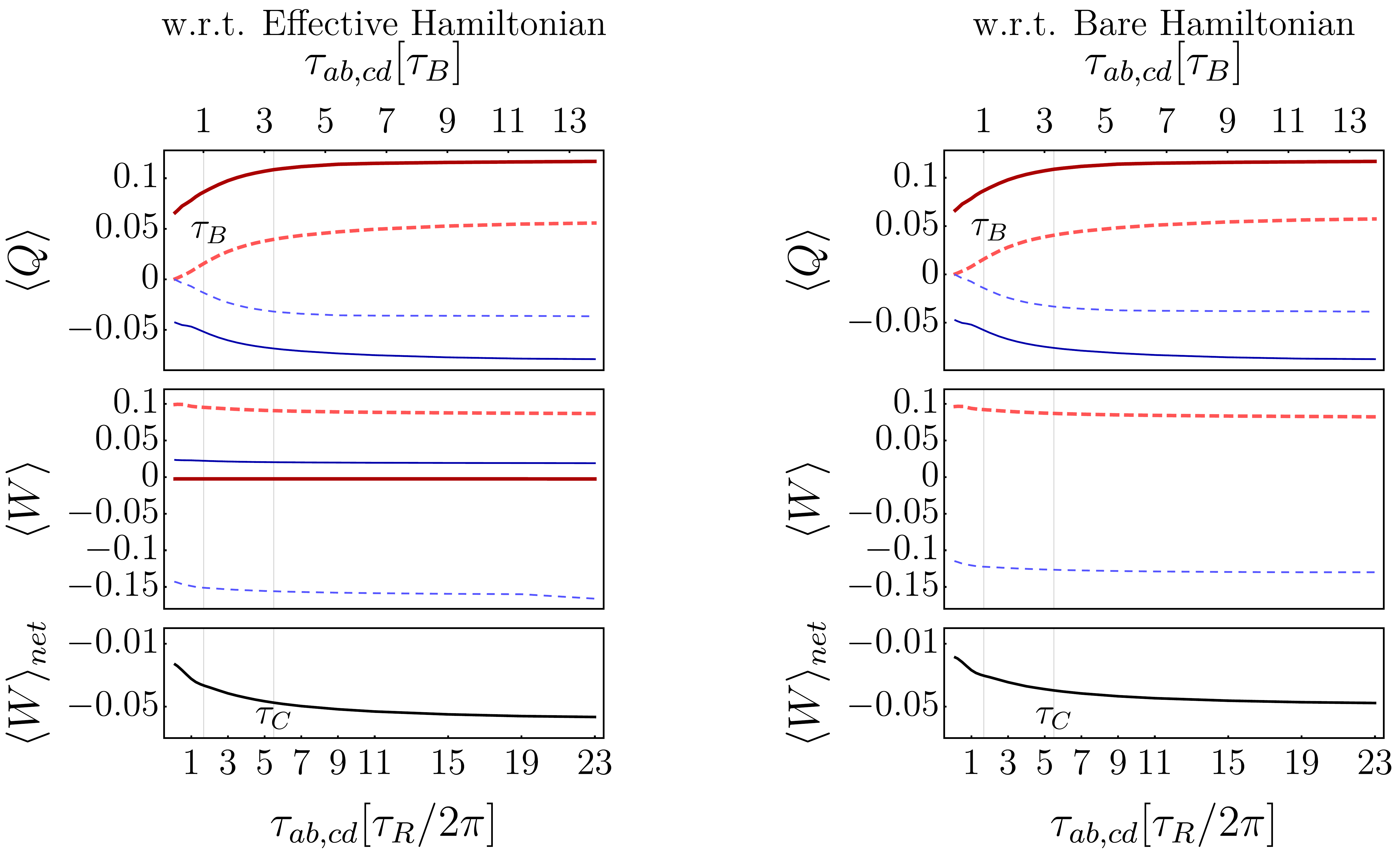}
\caption{The average heat and the average work during each strokes of the cycle and the net average work of a full cycle as a function of $\tau_{ab,cd}$. Here we set $f=2$ and $(g_c,g_h)=g_1$. The left panel shows the quantities calculated w.r.t. the effective Hamiltonian and the right panel with respect to the bare Hamiltonian. The thick dashed red curves correspond to $a\rightarrow b$ process, the thin solid blue curves to $b \rightarrow c$ process, the thin dashed blue to $c \rightarrow d$ process and the thick red curve to $d\rightarrow a$ process.}
\label{Fig_Energies}
\end{figure}
The average heat transferred, average work and the net average work are plotted in Fig.~\ref{Fig_Energies} considering the effective Hamiltonian $\hat{H}_{eff}$ in the left panel, and the bare Hamiltonian $\hat{H}_S$ in the right panel. Using the effective Hamiltonian to calculate the energy terms leads to some non-zero amount of average work for the isochoric strokes due to the time-dependent Lamb shifts. The corresponding terms are absent when we use the bare Hamiltonian as the drive is off during the isochoric strokes. Moreover, the average work in the expansion stroke is higher when we consider the effective Hamiltonian, which shows up also in the net extractable average work and consequently results into a higher efficiency in comparison to the case of using the bare Hamiltonian (see Fig.~\ref{Fig_Eff_Power}). This behavior is again due to the non-zero Lamb shift terms and the fact that by including them the effective frequency span of the WS is higher than the bare frequency span $\Delta \omega=\omega_2-\omega_1$, specifically for $c\rightarrow d$ process. The average net work approaches its adiabatic limit as we increase $\tau_{ab,cd}$ and decreases by speeding up the drive. One can see that the average heat transferred during the compression and expansion processes goes to zero as we decrease $\tau_{ab,cd}$, because the WS does not have enough time to exchange energy with the baths. Moreover, the heat transferred during the isochoric strokes reaches its non-zero minimum by approaching the diabatic limit (points $\bar{b}$ and $\bar{d}$ in Fi.g~\ref{Fig_diagSym}). 

Besides these asymptotic scenarios, we observe a dip in the heat transferred and the net average work at some values of $\tau_{ab,cd}$ coinciding with the peak in the efficiency. The dip especially indicates some extent of suppression of heat transferred to the cold bath. With a given amount of heat absorbed from the hot bath, if the WS dissipates less to the cold bath it means that the work done is higher and thereby the efficiency as well. This may suggest that a faster $a \rightarrow b'$ process in Fig.~\ref{Fig_diagSym} is in general beneficial, as the state at the end point $b'$ gets closer to the equilibrium state at the point $c$ and there would be less dissipated heat to the cold bath. However, the faster is $a \rightarrow b'$ process, the less amount of heat is absorbed from the hot bath which restricts the amount of extractable work too. Note that a similar situation also happens for the $c \rightarrow d'$ process considering the heat dissipated during the expansion and the heat absorbed during the thermalization $d' \rightarrow a$. Therefore, there must be some trade-off giving us the optimum efficiency in the intermediate situation. 

\begin{figure}
\centering
\includegraphics[width=0.7\linewidth]{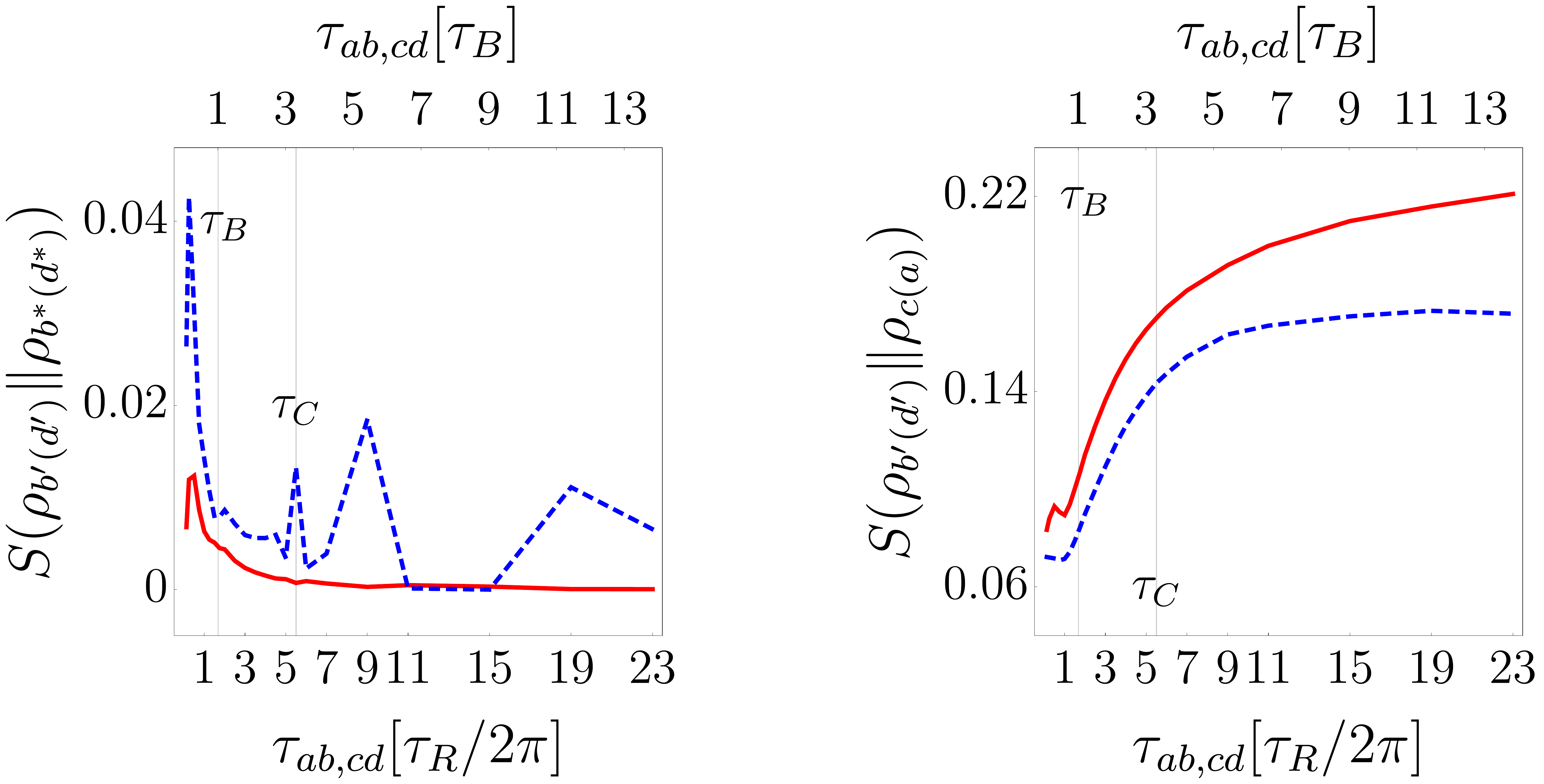}
\caption{Distance between the end point state $\hat{\rho}_{b'(d')}$ and the instantaneous steady state $\hat{\rho}_{b^*(d^*)}$ plotted in solid red(dashed blue) in the left panel, and the corresponding distance to the equilibrium state $\hat{\rho}_{c(d)}$ plotted using solid red(dashed blue)lines in the right panel. Here we set $f=2$ and $(g_c,g_h)=g_1$}
\label{Fig_distances}
\end{figure}

To shed some light on the facts discussed above, we consider the distance between the states at some end points in Fig.~\ref{Fig_diagSym}. First, the distance between $b'(d')$ and $b^*(d^*)$ allows us to figure how far we are from the instantaneous equilibrium states at the end of the compression and expansion processes. Second, the distance between the states at $b'(d')$ and $c(d)$ indicates how far the WS is from the thermal states at the end points of the isochoric strokes. To measure the distance between two states $\rho_1$ and $\rho_2$ we use the relative entropy between them defined by
\begin{equation}
S(\hat{\rho}_1\Vert \hat{\rho}_2)=\mathrm{tr}[\hat{\rho}_1\log(\hat{\rho}_1)]-\mathrm{tr}[\hat{\rho}_1\log(\hat{\rho}_2)].
\end{equation}
Looking at the left panel of Fig.~\ref{Fig_distances}, the distance between the end point state $\hat{\rho}_{b'(d')}$ and  the corresponding instantaneous asymptotic steady state $\hat{\rho}_{b*(d*)}$ decreases as we increase the driving duration, although there are some fluctuations in the process $c \rightarrow d'$. The interesting feature is that in the same time scale at which we observed the boost in the efficiency, the distance between $\hat{\rho}_{b'}$ and $\hat{\rho}_{b*}$ is close to its maximum, whereas, two states $\hat{\rho}_{d'}$ and $\hat{\rho}_{d}$ are rather close. The distance to the instantaneous asymptotic steady state $\hat{\rho}_{b*(d*)}$ is an indicator of irreversibility of the process, i.e. a large distance indicates higher irreversibility and smaller amount of extractable work. Looking at the right panel of Fig.~\ref{Fig_distances}, we realize a dip at $\tau_{ab,cd}\simeq \tau_B$. In addition, we observe that distance for the $a \rightarrow b'$ process is in general higher than $c \rightarrow d'$ process. 

An interesting feature about the results in Fig.~\ref{Fig_distances} is the asymmetry in the behaviors of $a\rightarrow b'$ and $c \rightarrow d'$ processes. This fact rises the question whether considering different speeds for the two processes has some non-trivial effects on the performance of the heat engine.
\begin{figure}
\centering
\includegraphics[width=0.5\linewidth]{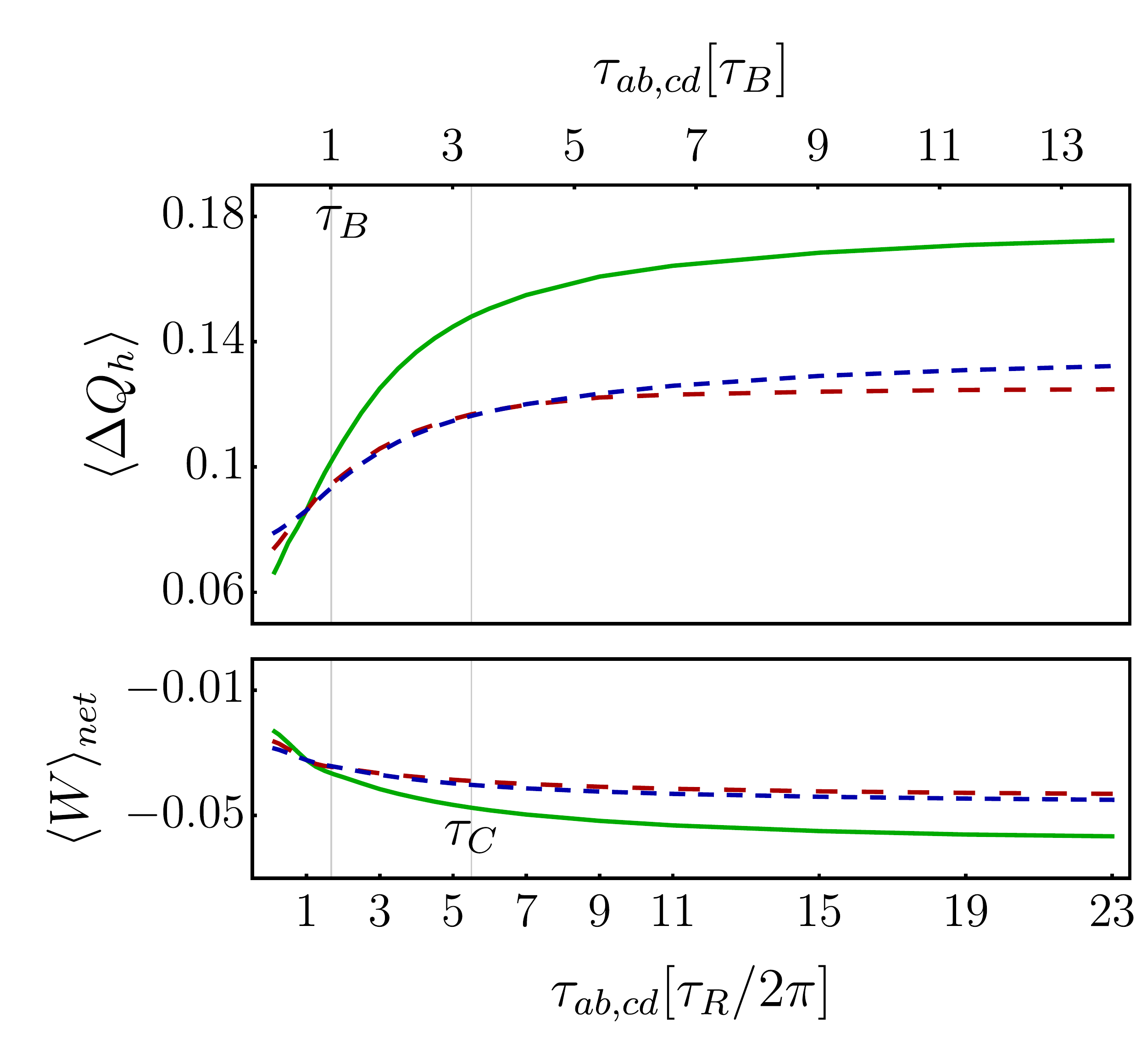}
\caption{The net average work and the net heat transferred into the WS as a function of $\tau_{ab,cd}$. The solid green curves correspond to the symmetric case with $\tau_{ab}=\tau_{cd}$, the large dashed red represents the asymmetric case (\textit{a}), and the small dashed blue to the asymmetric case (\textit{b}) discussed in the main text. Here we set $f=2$ and $(g_c,g_h)=g_1$.}
\label{Fig_work_heat_Asym_Sym}
\end{figure}
To make this point clear, we consider two different situations: (\textit{a}) setting $\tau_{ab}= \tau_D$ fixed while changing $\tau_{cd}$ and (\textit{b}) setting $\tau_{cd}=\tau_D$ fixed while varying $\tau_{ab}$. The efficiency and the average output power of these cases are plotted in Fig.~\ref{Fig_Eff_Power} using large dashing red and small dashing blue lines, respectively. Again the thick curves correspond to calculating the energies w.r.t. the effective Hamiltonian and the thin curves to the bare Hamiltonian. Interestingly, efficiency of the asymmetric cycles is always higher than the symmetric ones. However, superiority of the two asymmetric cases with respect to each other depends non-trivially on the time scale of the driving. Let us also examine the energetic of the asymmetric cycles in comparison to the symmetric ones depicted in Fig.~\ref{Fig_work_heat_Asym_Sym}. We note that the amount of net work is dependent on the total time of the expansion and compression process, $\tau_{tot}=\tau_{ab}+\tau_{cd}$, and in general decreases by decreasing $\tau_{tot}$ due to irreversibility. However, within the two asymmetric cycles with the same value of $\tau_{tot}$, we notice slightly different values for the net average work, indicating again the importance of the finite-time effects in the performance of the heat engines. 
\section{Conclusions}
In conclusion, we have studied the performances of a Stirling cycle when operated as a finite-time quantum heat engine. We first derived a non-Markovian master equation which allows us to study the dynamics of an open quantum system including counter rotating terms and avoids to make a clear distinction between the time scales of the system and of the environments. Thanks to this, we have been able to study the effect of the competing time scales, such as the typical time scale of the drive and the bath correlation/resonance time, on the performances of the heat engine. The main motivation of this work was to explore the performance of the heat engine operating in the non-adiabatic regime. Interestingly, we found that driving the WS at a time scale comparable to the resonance time of the bath, in addition to a boost in the output power, let us get an efficiency that is higher than the efficiency of the slow adiabatic cycle. One should note however that the net extractable work decreases by speeding up the cycle due to higher and higher degree of irreversibility. The other important finding in this work was the non-trivial dependency of the performance of the heat engine on the individual speed of the compression and expansion processes. Interestingly, one may achieve better performances by applying asymmetric compression and expansio speeds rather than a symmetric one. The latter opens new possibilities to optimize the performance of the quantum heat engines. An important aspect that is not covered in the current work is the reverse Stirling cycle working as a refrigerator. In \cite{Huang2014}, the authors have discussed that in general the revers cycle of a quantum Stirling heat engine might not be a refrigerator. As an outlook of our work, it is therefore interesting to explore how finite-time effects influence the  operating range of the quantum Stirling as a refrigerator and also its performance. In addition, our results motivate for further studies aiming at optimization of quantum thermodynamic cycles with finite-time driving protocols. 

\section*{Acknowledgement}
S. H. R. is thankful for the financial support from the Finnish Cultural Foundation and the Turku University Foundation. G. S. P. would like to acknowledge the FQXi project ``Exploring the fundamental limits set by thermodynamics in the quantum regime'' as well as the Academy of Finland through project no. 328193 and the  “Finnish
Center of Excellence in Quantum Technology QTF”, project no. 312296. S. M. and N. L. G. acknowledge financial support from the Academy of Finland Centre of Excellence program (Project no. 312058) and the Academy of Finland (Project no. 287750). N. L. G. acknowledges financial support from the Turku Collegium for Science and Medicine (TCSM).
\appendix
\section{Master equation}
\subsection{Numerical solution}
Here we briefly elaborate our numerical method to solve the master equation 
\begin{equation}\label{diss_appen}
\mathcal{L}_t[\hat{\rho}(t)]=-i[\hat{H}_S(t),\hat{\rho}(t)]+\int_0^t~d\tau\Phi(t-\tau)[\hat{\tilde{S}}(t,\tau)\hat{\rho}(t),\hat{S}(t)]+h.c.~.
\end{equation} 
The unitary propagator $\hat{U}(t,0)=\mathrm{T}\exp\left(-i\int _0^t d\tau~\hat{H}_S(\tau)\right)$ can be calculated numerically in a time interval $[0,t_{max}]$ by solving the Schrödinger equation
\begin{equation}
\frac{d}{dt}\hat{U}(t,0)=-i\hat{H}_S(t)\hat{U}(t,0).
\end{equation} 
Then owing to the divisibility of the unitary propagator we get
\begin{equation}
\hat{U}(t,\tau)=\hat{U}(t,0)\hat{U}(\tau,0)^{\dagger},~~0\leq\tau<t\leq t_{max}
\end{equation}
Inserting this solution in $\hat{\tilde{S}}(t,\tau)=\hat{U}(t,\tau)\hat{S}(\tau)\hat{U}(t,\tau)^{\dagger}$ and decomposing the operators with respect to Pauli operator basis $\{\hat{\sigma}_0=\hat{I},~\hat{\sigma}_x,~\hat{\sigma}_y,~\hat{\sigma}_z\}$ we get
\begin{equation}\label{sTildeSol}
\hat{\tilde{S}}(t,\tau)=\sum_{i}\tilde{s}_i(t,\tau)\hat{\sigma}_i.
\end{equation}
By introducing a similar decomposition for the operator $\hat{S}(t)$ given by $\hat{S}(t)=\Sigma_j\lambda_j(t)\hat{\sigma}_j$, the second term on the r.h.s. of Eq.~(\ref{diss_appen}) will be rewritten as
\begin{equation}
\sum_{i,j}R_{ij}(t)[\hat{\sigma}_i\hat{\rho}(t),\hat{\sigma}_j(t)]+h.c.,
\end{equation}
with the time-dependent rates
\begin{equation}
R_{ij}(t)=\lambda_j(t)\int_0^t~d\tau\Phi(t-\tau)\tilde{s}_i(t,\tau).
\end{equation}

\subsection{ME decomposed with respect to the instantaneous energy basis of the open quantum system}
Consider an instantaneous eigenvector of $\hat{H}_S(t)$ denoted by $\ket{\epsilon_i(t)}$ corresponding to the instantaneous energy eigenvalue $\epsilon_i(t)$. By defining $\hat{E}_{nm}(t)=\ket{\epsilon_n(t)}\bra{\epsilon_m(t)}$,
one can decompose $\hat{S}$ and $\hat{\tilde{S}}$ as
\begin{eqnarray}
&&\hat{\tilde{S}}(t,\tau)=\sum_{n,m}\tilde{\xi}_{nm}(t,\tau)\hat{E}_{nm}(t),
\\
&&\hat{S}(t)=\lambda(t)\sum_{n,m}\eta_{nm}(t)\hat{E}_{nm}(t).
\end{eqnarray}
By inserting these expressions in Eq.~(\ref{diss_appen}), the second term on the r.h.s. takes a form given by
\begin{eqnarray}\label{dissInst}
\sum_{n,m}\sum_{r,s}\Big\{&& R_{nm,rs}^{(\downarrow)}(t)\big[\hat{E}_{nm}(t)\hat{\rho}(t)\hat{E}_{rs}(t)-\hat{E}_{rs}(t)\hat{E}_{nm}(t)\hat{\rho}(t)\big]
\\
&&+R_{nm,rs}^{(\uparrow)}(t)\big[\hat{E}_{rs}(t)\hat{\rho}(t)\hat{E}_{nm}(t)-\hat{\rho}(t)\hat{E}_{nm}(t)\hat{E}_{rs}(t)\big]\Big\},\nonumber
\end{eqnarray}
with
\begin{eqnarray}
R_{nm,rs}^{(\downarrow)}(t)=\lambda(t)\int_0^t~d\tau\Phi(t-\tau)\tilde{\xi}_{nm}(t,\tau)\eta_{nm}(t),
\\
R_{nm,rs}^{(\uparrow)}(t)=\lambda(t)\int_0^t~d\tau\Phi(t-\tau)^{*}\tilde{\xi}_{nm}(t,\tau)\eta_{nm}(t).
\end{eqnarray}
One can further arrange Eq.~(\ref{dissInst}) into rotating (R) and counter-rotating (CR) parts with respect to the instantaneous energy basis. The rotating part takes the form
\begin{eqnarray}
\mathcal{L}^{(R)}_t[\hat{\rho}(t)]=\sum_{n\neq m}\Big\{&&R_{nm,mn}^{(\downarrow)}(t)\big[\hat{E}_{nm}(t)\hat{\rho}(t)\hat{E}_{mn}(t)-\hat{E}_{mn}(t)\hat{E}_{nm}(t)\hat{\rho}(t)\big]
\\
&&+R_{nm,mn}^{(\uparrow)}(t)\big[\hat{E}_{mn}(t)\hat{\rho}(t)\hat{E}_{nm}(t)-\hat{\rho}(t)\hat{E}_{nm}(t)\hat{E}_{mn}(t)\big]\Big\},\nonumber
\end{eqnarray}
while the counter-rotating part $\mathcal{L}_t^{(CR)}$ includes all the remaining terms.

We focus now on the specific model considered in this paper given by the Hamiltonian in Eq.~(\ref{Hamiltonian}), and  instantaneous energy basis labeled by $\ket{\epsilon_e(t)}$ and $\ket{\epsilon_g(t)}$. Having $\hat{S}(t)=\lambda(t)\hat{\sigma}_y$, and the expression for the energy basis given in Eq.~(\ref{basis_e}) and Eq.~(\ref{basis_g}), we get
\begin{equation}\label{sDecom}
\hat{S}(t)=\lambda(t)\left(-i \hat{E}_{eg}(t)+i \hat{E}_{ge}(t)\right).
\end{equation}
Considering the numerical solution in Eq.~(\ref{sTildeSol}), a decomposition for $\hat{\tilde{S}}=\sum_{n,m=e,g}\tilde{s}_{nm}(t,\tau)\hat{E}_{nm}(t)$ in the energy basis is given by 
\begin{eqnarray}\label{sTildeDecom}
\hat{\tilde{S}}(t,\tau)&=
\\
&\left(\tilde{s}_0+\frac{\tilde{s}_x\Delta+\tilde{s}_zq(t)}{\sqrt{q(t)^2+\Delta^2}}\right)\hat{E}_{ee}(t)+\left(\tilde{s}_0-\frac{\tilde{s}_x\Delta+\tilde{s}_zq(t)}{\sqrt{q(t)^2+\Delta^2}}\right)\hat{E}_{gg}(t)\nonumber
\\
&+\left(-i\tilde{s}_y+\frac{\tilde{s}_x q(t)-\tilde{s}_z\Delta}{\sqrt{q(t)^2+\Delta^2}}\right)\hat{E}_{eg}(t)+\left(i\tilde{s}_y+\frac{\tilde{s}_x q(t)-\tilde{s}_z\Delta}{\sqrt{q(t)^2+\Delta^2}}\right)\hat{E}_{ge}(t),\nonumber
\end{eqnarray}
where $\tilde{s}_{i}\equiv\tilde{s}_i(t,\tau)$. Note that one has $R_{eg,ge}^{(\downarrow)}(t)=R_{ge,eg}^{(\uparrow)}(t)^*$ and $R_{ge,eg}^{(\downarrow)}(t)=R_{eg,ge}^{(\uparrow)}(t)^*$. Moreover, since $\tilde{s}_0(t,\tau)\equiv0$ and all other $\tilde{s}_i$ are real valued, we also have $R_{ee,ge}^{(\downarrow)}(t)=-R_{gg,eg}^{(\downarrow)}(t)^*$ and $R_{ee,eg}^{(\uparrow)}(t)=-R_{gg,ge}^{(\uparrow)}(t)^*$. Accordingly, the rotating part $\mathcal{L}_t^{(R)}$ reads
\begin{eqnarray}
\mathcal{L}^{(R)}_t[\hat{\rho}(t)]=&-i\left [\delta^{(R)}_L(t)\hat{H}_S(t),\hat{\rho}(t)\right]
\\
&~+\gamma^ {(\downarrow)}(t)\Big[\hat{L}(t)\hat{\rho}(t)\hat{L}^{\dagger}(t)-\frac{1}{2}\{\hat{L}^{\dagger}(t)\hat{L}(t),\hat{\rho}(t)\}\Big] \nonumber
\\
&~~+\gamma^{(\uparrow)}(t)\Big[\hat{L}^{\dagger}(t)\hat{\rho}(t)\hat{L}(t)-\frac{1}{2}\{\hat{L}(t)\hat{L}^{\dagger}(t),\hat{\rho}(t)\}\Big],\nonumber
\end{eqnarray}
where $\hat{L}(t)=\hat{E}_{ge}(t)$. The explicit expressions for the rates are $\gamma^{(\downarrow)}(t)=2\mathrm{Re}[R_{ge,eg}^{(\downarrow)}(t)]$, $\gamma^{(\uparrow)}(t)=2\mathrm{Re}[R_{ge,eg}^{(\uparrow)}(t)^*]$, and $\delta^{(R)}_L(t)=(\mathrm{Im}[R_{ge,eg}^{(\downarrow)}(t)]+\mathrm{Im}[R_{ge,eg}^{(\uparrow)}(t)^*])/2$. In addition the counter-rotating Lamb shift is given by $\delta^{(CR)}_L(t)=\mathrm{Im}[R_{ee,eg}^{(\downarrow)}(t)]$. However, the expression for the counter-rotating dissipator is so lengthy that does not fit here. 
\section{Analytic considerations for the asymptotic Stirling cycles}
Apart from the dynamical approach of the paper, we provide some analytic analysis of the energy transferred for some asymptotic cases. Assume that at the end of the isochoric strokes $b \rightarrow c$ and $d \rightarrow a$ the TLS relaxes to the corresponding thermal states
\begin{equation}
\hat{\rho}_{c}=\frac{e^{-\beta_c \hat{H}_S(t_c)}}{\mathrm{tr}[e^{-\beta_c \hat{H}_S(t_c)}]},~~~\hat{\rho}_{a}=\frac{e^{-\beta_h \hat{H}_S(t_a)}}{\mathrm{tr}[e^{-\beta_h \hat{H}_S(t_a)}]}
\end{equation}  
With this assumption and by considering the bare Hamiltonian of the TLS, one can analyze the energetic of the quantum Stirling cycle with regards to the four asymptotic cases listed below.
\subsection*{The (ss) cycle: ideally slow compression and slow expansion}
In this extreme, both the compression and the expansion processes are ideally slow, i.e. the qubit follows a trajectory on which it is always at thermal equilibrium with the bath. The heat transfer during the four strokes then is calculated by \cite{Huang2014}($\hbar=1$, $k_B=1$) 
\begin{eqnarray}
&&\langle Q_{ab}\rangle=\beta_h\int_a^b dS=\beta_h[S(\hat{\rho}_{t_b})-S(\hat{\rho}_{t_a})],
\\
&&\langle Q_{bc}\rangle=\mathrm{Tr}[\hat{H}_S(t_c)\hat{\rho}_{t_c}]-\mathrm{Tr}[\hat{H}_S(t_b)\hat{\rho}_{t_b}],
\\
&&\langle Q_{cd}\rangle=\beta_h\int_c^d dS=\beta_h[S(\hat{\rho}_{t_d})-S(\hat{\rho}_{t_c})],
\\
&&\langle Q_{da}\rangle=\mathrm{Tr}[\hat{H}_S(t_d)\hat{\rho}_{t_d}]-\mathrm{Tr}[\hat{H}_S(t_a)\hat{\rho}_{t_a}],
\end{eqnarray}
where $S(\hat{\rho})$ is the von Neumann entropy of a given state $\hat{\rho}$. For a two-level system with the level populations $\rho_{ee}$ and $\rho_{gg}$ one has
\begin{equation}
S(\hat{\rho})=-(\rho_{ee}\log[\rho_{ee}]+\rho_{gg}\log[\rho_{gg}]).
\end{equation}
Then according to the first law of thermodynamics we get the net average work done on the qubit by $\langle W\rangle_{net}=- (\langle Q_{ab}\rangle+\langle Q_{bc}\rangle+\langle Q_{cd}\rangle+\langle Q_{da}\rangle)$.
\subsection*{The (fs) cycle: ideally fast compression and slow expansion}
In this extreme, $a \rightarrow b$ process is done in a finite but very fast time scale, such that the process is diabatic. For a sufficiency fast process, TLS does not have time to exchange heat with the hot bath and $\langle Q_{ab}\rangle=0$. Nonetheless, there is some non-zero average work done on the TLS that can be obtained by the change in its internal energy. Since the process is diabatic, state of the TLS remains at its initial configuration at time $t_a$, therefore
\begin{equation}
\langle W_{ab}\rangle=\mathrm{tr}\left[(\hat{H}_S(t_b)-\hat{H}_S(t_a))\hat{\rho}_{t_a}\right].
\end{equation}
The remaining energy terms can be calculated similar to the $(ss)$ case.  
\subsection*{The (sf) cycle: ideally slow compression and fast expansion}
This is the opposite situation of the $(fs)$ cycle, such that $\langle Q_{cd}\rangle=0$ and
\begin{equation}
\langle W_{cd}\rangle=\mathrm{tr}\left[(\hat{H}_S(t_d)-\hat{H}_S(t_c))\hat{\rho}_{t_c}\right].
\end{equation}
\subsection*{The (ff) cycle: ideally fast compression and fast expansion}
Finally when both the processes are diabatic, one has $\langle Q_{ab}\rangle=\langle Q_{cd}\rangle=0$ and the amounts of average work can be obtained as discuss in the two previous cases.
\section*{Bibliography}
\bibliographystyle{iopart-num}
\bibliography{QSC_Draft_V4}

\end{document}